\documentclass[12pt]{article}
\newcommand{\bra}[1]{\langle#1\vert}
\newcommand{\ket}[1]{\vert#1\rangle}
\newcommand{\braket}[2]{\langle#1\vert#2\rangle}
\newcommand{\ketbra}[2]{\vert#1\rangle\langle#2\vert}
\newcommand\sandwich[3]{\langle#1\vert#2\vert#3\rangle}
\newcommand{\be}{\begin{equation}}
\newcommand{\ee}{\end{equation}}
\newcommand{\bea}{\begin{eqnarray}}
\newcommand{\eea}{\end{eqnarray}}
\newcommand{\bq}{\begin{quote}}
\newcommand{\eq}{\end{quote}}
\newcommand{\e}{\varepsilon}
\newcommand{\cS}{{\cal S}}
\newcommand{\E}{{\cal E}}
\newcommand{\A}{{\cal A}}
\newcommand{\cP}{{\cal P}}
\newcommand{\W}{{\cal W}}
\newcommand{\R}{{\cal R}}

\begin{document}
\setlength{\baselineskip}{16.5pt}
\title{PROBABILITIES FROM ENVARIANCE?}
\author{Ulrich Mohrhoff\\
Sri Aurobindo International Centre of Education\\
Pondicherry 605002 India\\
\normalsize\tt ujm@auromail.net}
\date{}
\maketitle
\begin{abstract}
\noindent Zurek claims to have derived Born's rule noncircularly in the context of 
an ontological no-collapse interpretation of quantum states, without any ``{\it 
deus ex machina\/} imposition of the symptoms of classicality.'' After a brief 
review of Zurek's derivation it is argued that this claim is exaggerated if not 
wholly unjustified. In order to demonstrate that Born's rule arises noncircularly 
from deterministically evolving quantum states, it is not sufficient to assume that 
quantum states are somehow associated with probabilities and then prove that 
these probabilities are given by Born's rule. One has to show how irreducible 
probabilities can arise in the context of an ontological no-collapse interpretation 
of quantum states. It is argued that the reason why all attempts to do this have 
so far failed is that quantum states are fundamentally algorithms for computing 
correlations between possible measurement outcomes, rather than evolving 
ontological states.

\vspace{6pt}\noindent {\it Keywords\/}: Born's rule; envariance; 
interpretation of quantum mechanics; probability.
\setlength{\baselineskip}{14pt}
\end{abstract}

\section{\large Introduction}
In any metaphysical framework that treats quantum states as deterministically 
evolving ontological states, such as Everett's many-worlds interpretation, Born's 
rule has to be postulated. Repeated attempts to derive Born's rule within the 
many-worlds framework have proved circular, as was noted even by its 
proponents. (For references see~\cite{ZurekRMP}.) In the context of 
environment-induced decoherence, Born's rule emerges 
naturally~\cite{ZurekRG} but decoherence is based on reduced density matrices, 
and the partial tracing that leads to reduced density matrices is predicated on 
Born's rule. No noncircular derivation of Born's rule has ever been put forward in 
the context of an ontological no-collapse interpretation of quantum states.

Recently, however, Wojciech H. Zurek has claimed to have shown how Born's rule 
arises noncircularly ``in a purely quantum setting, i.e., without appeals to 
`collapse,' `measurement,' or any other such {\it deus ex machina\/} imposition 
of the symptoms of classicality that violate the unitary spirit of quantum theory'' 
\cite{ZurekRMP,ZurekEnv,ZurekDarwin}. After a brief review of Zurek's 
derivation in Sec.~2, it is argued that this claim is exaggerated if not wholly 
unjustified. In order to demonstrate that Born's rule arises noncircularly from 
deterministically evolving ontological quantum states (DEOQS), it is not sufficient 
to assume that quantum states are somehow associated with probabilities and 
then prove that these probabilities are given by Born's rule, as Zurek has done.

Section 3 offers an instructive attempt to complete Zurek's derivation by 
demonstrating how irreducible probabilities---probabilities associated not with 
averaged states but with microstates of individual systems---can arise in the 
context of an ontological no-collapse interpretation of quantum states. The 
attempt fails because it tacitly relies on decoherence and thus makes implicit use 
of Born's rule. It is argued in Sec.~4 that this is not a failing of only this particular 
attempt but a failing of the notion that quantum states are evolving ontological 
states, rather than being fundamentally algorithms for computing correlations 
between possible measurement outcomes. It is further argued that Gleason's 
theorem~\cite{Gleason} and more recent Gleason-like derivations of Born's rule 
\cite{Fuchs,Busch,Cavesetal} offer a deeper understanding of the axioms of 
standard quantum theory than Zurek's does. (See also the comment by 
Schlosshauer and Fine~\cite{SchlossFine} on Zurek's derivation of Born's rule.)

\section{\large Zurek's derivation of Born's rule}
The following assumptions are made:
\begin{enumerate}
\item[(i)] The universe consists of systems.
\item[(ii)] The states of a system $\cS$ are (associated with) the normalized 
elements $\ket\psi$ of a Hilbert space that describes~$\cS$.
\item[(iii)] A composite system is described by the tensor product of the Hilbert 
spaces of the constituent systems.
\item[(iv)] States evolve according to $i\hbar\ket{\dot \psi}=H\ket\psi$ 
where $H$ is Hermitean.
\end{enumerate}
{\it Envariance\/} (environment--assisted invariance) is defined as follows: If 
for a vector $\ket{\psi_{\cS\&\E}}$ associated with the composite 
system~$\cS\&\E$, where $\cal E$ is a dynamically decoupled environment of 
$\cS$, and for a transformation $U_\cS = u_\cS \otimes {\bf 1}_\E$ there 
exists a transformation $U_\E = {\bf 1}_\cS \otimes u_\E$ such that
\be
U_\E U_\cS\ket{\psi_{\cS\&\E}} = \ket{\psi_{\cS\&\E}}
\ee
then $\ket{\psi_{\cS\&\E}}$ is {\it envariant\/} under $u_\cS$.

Suppose that
\be
\sum_{k=1}^N a_k\ket{s_k}\otimes\ket{\e_k}\label{sch}
\ee
is a biorthonormal (Schmidt) decomposition of $\ket{\psi_{\cS\&\E}}$, and 
that
\be
u_\cS\ket{s_k}=e^{i\phi_k}\ket{s_k},\quad k=1,\dots,N.
\ee
For any set of coefficients~$a_k$ and any set of integers $l_k$ the effect of 
$u_\cS$ on $\ket{\psi_{\cS\&\E}}$ can be undone by
\be
u_\E\ket{\e_k}=e^{-i(\phi_k+2\pi l_k)}\ket{\e_k}.
\ee
Thus $\ket{\psi_{\cS\&\E}}$ is envariant under $u_\cS$. Since the properties 
of $\cS\&\E$, as well as the respective properties of $\cS$ and~$\E$, are fully 
determined by $\ket{\psi_{\cS\&\E}}$, no feature of $\ket{\psi_{\cS\&\E}}$ 
that is affected by $u_\cS$ can represent a property that is possessed by $\cS$ 
while the joint state of $\cS\&\E$ is $\ket{\psi_{\cS\&\E}}$. It follows that 
the phases of the coefficients~$a_k$ in a biorthogonal decomposition of 
$\ket{\psi_{\cS\&\E}}$ cannot represent properties of (or contain information 
about)~$\cS$.

If any two of the coefficients in the decomposition~(\ref{sch}) have equal 
norms---say, $a_1=|a|e^{-i\phi_1}$ and $a_2=|a|e^{-i\phi_2}$---then 
$\ket{\psi_{\cS\&\E}}$ is also envariant under the {\it swap\/}
\be
u_\cS(1{\leftrightarrow}2)=
e^{i\phi_{12}}\ketbra{s_1}{s_2}+e^{-i\phi_{12}}\ketbra{s_2}{s_1},
\ee
which can be undone by the ``counterswap''
\be
u_\E(1{\leftrightarrow}2)=
e^{-i(\phi_{12}+\phi_1-\phi_2+2\pi l_{12})}\ketbra{\e_1}{\e_2}
+e^{i(\phi_{12}+\phi_1-\phi_2+2\pi l_{12})}\ketbra{\e_2}{\e_1}.
\ee
It follows that the information about~$\cS$ contained in 
$\ket{\psi_{\cS\&\E}}$ is envariant under $u_\cS(1{\leftrightarrow}2)$. This 
leads Zurek ``to conclude immediately that the probabilities for any two 
swappable $\ket{s_k}$ are equal''~\cite{ZurekEnv}. He further concludes that 
if all of the coefficients in the decomposition~(\ref{sch}) have equal norms---in 
which case $\ket{\psi_{\cS\&\E}}$ is invariant under any re-labeling of the 
states~$\ket{s_k}$---the probability $p(s_k)$ must be equal to~$1/N$ for 
all~$k$. Observers who know the state of $\cS\&\E$ to be 
$\ket{\psi_{\cS\&\E}}$ with $a_k=e^{-i\phi_k}/\sqrt{N}$ are ``provably 
ignorant'' of (the state or the properties of)~$\cS$. They will therefore 
``conclude that the probabilities of all the envariantly swappable outcomes must 
be the same''~\cite{ZurekRMP}.

To complete his derivation of Born's rule, Zurek considers the case in which the 
coefficients $a_k$ in the decomposition (\ref{sch}) are proportional to 
$\sqrt{m_k}$ with natural numbers~$m_k$, so that the norms $|a_k|^2$ are 
commensurate. (The assumption that all coefficients have the same phase is 
warranted by the fact that, as shown, the phases of the Schmidt coefficients are 
irrelevant as far as the properties of~$\cS$ concerned.) He then extracts from 
$\E$ a ``counterweight''~$\A$ such that
\be
\ket{\psi_{\A\&\cS\&\E}}\propto
\sum_{k=1}^N\sqrt{m_k}\ket{A_k}\otimes\ket{s_k}\otimes\ket{\e_k}.
\ee
By ``increasing the resolution of~$\A$'',
\be
\ket {A_k}=\sum_{j_k=1}^{m_k}\ket{a_{j_k}}/\sqrt{m_k},
\ee
and letting $\A$ interact with $\E$ so that
\be
\ket{a_{j_k}}\otimes\ket{\e_k}\rightarrow\ket{a_{j_k}}\otimes\ket{e_{j_k
}}
\ee
with orthonormal vectors $\ket{e_{j_k}}$, he reduces this case to the case of 
coefficients with equal norms. The upshot: Born's rule $p(s_k)=m_k/M$, where 
$M=\sum_{k=1}^N m_k$. The generalization to coefficients with 
incommensurate norms is straightforward.

\section{\large Whence the probabilities?}
As said, in order to demonstrate that Born's rule arises noncircularly from\break
DEOQS, it is not sufficient to assume that quantum states are somehow 
associated with probabilities and then prove that these probabilities are given by 
Born's rule. What is thereby proved is that {\it if\/} quantum states are 
associated with probabilities then Born's rule holds. But how do quantum states 
come to be associated with probabilities? As long as this questions remains 
unanswered, one has not elucidated the origin of probabilities in quantum 
physics, as Zurek claims to have done~\cite{ZurekEnv}.

Zurek abruptly concludes that ``the probabilities for any two swappable 
$\ket{s_k}$ are equal.'' If these states are associated with probabilities then it 
is certainly the case that ``swappable'' implies ``equiprobable.'' But the idea 
that states are associated with probabilities comes out of the blue, without 
justification. What envariance under $u_\cS(1{\leftrightarrow}2)$ does imply is 
a symmetry with regard to the properties $s_1$ and~$s_2$. If these properties 
are mutually exclusive, $\cS$~cannot simultaneously possess both, but it may 
possess neither of them: the propositions ``$\cS$~is/has~$s_i$'' ($i{=}1,2$) 
may both be false, or else they may both lack truth values (i.e., the physical 
situation may be such that these propositions are neither true nor false but 
meaningless).

Can Zurek's incomplete derivation of Born's rule from DEOQS be completed by 
demonstrating how irreducible probabilities---probabilities associated not with 
averaged states but with microstates of individual systems---can arise in the 
context of an ontological no-collapse interpretation of quantum states? Here is 
how one might attempt to do this. If all of the coefficients in the 
decomposition~(\ref{sch}) have equal norms, and if the properties~$s_k$ are 
mutually exclusive and jointly exhaustive (i.e., if the normalized $\ket{s_k}$ 
form an orthonormal basis) then the propositions ``$\cS$~is/has~$s_k$'' 
cannot all be false. We already know that they cannot all be true, and that if they 
are in possession of truth values, they must be in possession of identical truth 
values. We conclude that they cannot be in possession of truth values. If all 
Schmidt coefficients have equal norms and the~$\ket{s_k}$ form an 
orthonormal basis, these propositions are neither true nor false. But if predicative 
propositions can lack truth values, a criterion has to exist: under what conditions 
does such a proposition possess a truth value?

In the spirit of Zurek's ``existential interpretation'' \cite{ZurekRMP,ZurekRG} 
(in the context of which Zurek's claims ought to be evaluated, if only to be fair) 
we may require the existence of a record. A truth value exists if and only if one is 
indicated by, or inferable from, a predictably evolving pointer state---a record. 
Once we have reached this point, we can with sufficient justification assign 
probabilities to all possible value-indicating states or records (``measurement 
outcomes'') inasmuch as we cannot predict the actual outcome. (If we could, we 
could assign truth values to the propositions ``$\cS$~is/has~$s_k$'' in advance 
of the outcome.) And if all coefficients in the decomposition~(\ref{sch}) have 
equal norms, we can invoke the principle of indifference and deduce from the 
nonexistence of truth values that all of those propositions are equally likely to 
come out true in the event that a record of their truth values is created.

Is this attempt to complete Zurek's derivation of Born's rule noncircular? I don't 
think so. Probabilities are associated with possibilities, and the relevant 
possibilities are possible measurement outcomes, or outcome-indicating states, 
or records. How do records enter the picture? By Zurek's account, through 
correlations. The recorded properties are those that are entangled with the most 
predictable states of the environment or of parts thereof. And the most 
predictable states are the most abundantly monitored ones---the ones entangled 
with the largest number of subsystems of the environment. The information 
recorded in them is so abundantly replicated that it is for all practical purposes 
indelible. And why are the most abundantly monitored states the most 
predictable ones? The chief ingredient in Zurek's answer to this question is 
decoherence, and decoherence involves in an essential way reduced density 
matrices, partial tracing, and thus, ultimately, an implicit use of Born's rule.

Is there any other way to complete Zurek's derivation noncircularly? I don't think 
so. The symbols $s_k$ and $\e_k$ in (\ref{sch}) stand for possible properties of 
$\cS$ and~$\E$. Predicative propositions are needed to relate these properties 
to their respective systems. Since these propositions are not necessarily in 
possession of truth values, a necessary condition for the existence of a truth 
value is required. There is broad consensus that, in the context of standard 
quantum mechanics (unadulterated with, e.g., Bohmian 
trajectories~\cite{Bohm} or nonlinear modifications of the ``dynamical'' 
equations~\cite{GRW}) this involves environment-induced decoherence in an 
essential way.

There are of course ways to cloud the issue, such as Zurek's 
suggestion~\cite{ZurekEnv} that the uniqueness of a Schmidt decomposition (in 
case none of the coefficients have equal norms) is sufficient for the existence of a 
unique set of {\it pointer\/} states. States selected by a unique Schmidt 
decomposition are not necessarily pointer states. To be indicators of 
measurement outcomes, states must {\it retain\/} information about outcomes, 
and decoherence arguments are essential for establishing which states are 
capable of retaining such information. Information about the (relative) values of 
an observable (relative as in ``relative state''~\cite{ZurekRMP}) can be 
contained (via a unique Schmidt decomposition) in a system's nonlocal properties 
(or in the values of a nonlocal observable) in which case the system or observable 
is exceedingly unlikely to retain this information, owing to environment-induced 
decoherence.

\section{\large Moral}
By simply assuming that quantum states are somehow associated with 
probabilities, Zurek has glossed over the fact that the reason why quantum states 
are associated with probabilities involves decoherence and hence Born's rule. 
Probabilities are associated with possibilities, and the relevant possibilities are 
measurement outcomes or records thereof. I don't see a way of introducing 
probabilities without introducing measurement outcomes, no matter what other 
names we invent for them. My failed attempt to complete Zurek's derivation 
noncircularly is a case in point. To be able to justify the introduction of 
probabilities, I had to assume that the properties~$s_k$ are mutually exclusive 
and jointly exhaustive. This assumption owes its meaning to measurements. The 
physical meaning of ``mutually exclusive'' (as against its mathematical 
implementation through orthogonality) is that {\it if\/} the propositions 
``$\cS$~is/has~$s_k$'' are in possession of measured truth values then the 
truth of one implies the falsity of the others. The physical meaning of ``jointly 
exhaustive'' is that {\it if\/} these propositions have measured truth values then 
at least one of them is true. Zurek's claim to have explained ``how Born's rule 
arises\dots without appeals to `collapse,' `measurement,' or any other such {\it 
deus ex machina\/} imposition of the symptoms of 
classicality''~\cite{ZurekEnv} is therefore unjustified.

Zurek emphasizes that, in contrast with other derivations of Born's rule, his 
``relies on the most quantum of foundations---the incompatibility of the 
knowledge about the whole and about the parts, mandated by entanglement.'' 
``Entanglement,'' too, owes its meaning to measurements. Its physical meaning 
is that one can measure (at any rate, obtain information about) the value of an 
observable~$V$ by measuring the value of another observable~$W$. Besides, 
what has knowledge to do with quantum foundations in the context of an {\it 
ontological\/} interpretation of quantum states? The statement that knowledge 
about the whole implies ignorance of the parts has two parts, one that conveys a 
lack of factuality and one that is counterfactual. Saying that a system composed 
of two spin-1/2 particles is in the singlet state is the same as saying (i)~that the 
components of the individual spins {\it lack\/} values, and (ii)~that if a spin 
component of each particle {\it were\/} measured then the results would be 
correlated as specified by the singlet state. There is therefore nothing factual 
about this ``state,'' nothing that would warrant an ontological interpretation. If 
there {\it is\/} an ontological state that is responsible for the correlations 
encapsulated by the singlet state, it is {\it not\/} the singlet state, and that's 
{\it all\/} we know about it.

The rather mystical-sounding statement that knowledge about the whole implies 
ignorance of the parts is thus largely a statement about correlated probability 
distributions over measurement outcomes. Given its implicit reference to 
probabilities, it does not elucidate the ``origin of 
probabilities''~\cite{ZurekRMP} but rather shows that probabilities are present 
from the start, however cleverly they may be concealed by mystical language. If 
the quantum formalism is (i)~the fundamental theoretical framework of physics 
and (ii)~fundamentally what somehow it obviously is (namely, a probability 
algorithm) then there is nothing "more fundamental"---a contradiction in terms 
since what is ``less fundamental'' just isn't fundamental---from which 
probabilities could emerge.

To my mind, the conclusion to be drawn from the past failures (including Zurek's) 
to derive probabilities noncircularly from DEOQS, is that quantum states {\it 
are\/} probability measures and should not be construed as evolving ontological 
states. Theorists ought to think of them the way experimentalists use them, 
namely as algorithms for computing the probabilities of possible measurement 
outcomes on the basis of actual measurement outcomes. No end of 
pseudophysics~\cite{Pseudo} is generated by the notion that quantum states 
are evolving ontological states. Then one is presented with two modes of 
evolution, one unitary and one projective, and a host of ensuing pseudoproblems. 
The way to get rid of these pseudoproblems is not to reject {\it one\/} mode of 
evolution but to reject them both. The lawlike features of the world are 
encapsulated in correlations between value-indicating events, not in an evolving 
ontological state. (The time dependence of quantum states is a dependence on 
the time of a measurement, relative to the time of another measurement, not the 
dependence of a state of affairs that persists and evolves.)

As long as we believe in DEOQS, the assumptions listed at the beginning of 
Sec.~2 are nothing short of baffling. They can be rendered comparatively 
transparent by preceding them with another assumption:
\begin{itemize}
\item The mathematical formalism of quantum mechanics is an algorithm for 
assigning probabilities to possible measurement outcomes on the basis of actual 
measurement outcomes.
\end{itemize}
As the following will show, this assumption elucidates assumptions (i)--(iv) {\it 
provided\/} that one does not invoke assumption~(iii) to derive Born's rule, as 
Zurek does.

\begin{enumerate}
\item[(i)] Why does the universe consist of systems? Because our fundamental 
theoretical framework is an algorithm that correlates possible outcomes of 
measurements that may be performed either on the same system at different 
times or on different systems in spacelike relation.

\smallskip
\item[(ii)] Why are the states of $\cS$ normalized elements $\ket\psi$ of a 
Hilbert space?

\smallskip
(a) {\it Hilbert space\/} because we want nontrivial probabilities (probabilities 
greater than~0 and less than~1) to be assigned to the possible outcomes of 
maximal measurements (measurements yielding the greatest possible amount of 
information) even if they are assigned on the basis of outcomes of maximal 
measurements. {\it Nontrivial probabilities\/} because otherwise we would have 
no reason to treat quantum states as probability measures.

[Classically, a property is represented by a subset~$\W$ of some phase space, 
and the probability measure determined by a maximal measurement is 
represented by a (0-dimensional) point~$\cP$ in that space. All probabilities are 
therefore trivial: 1~if $\cP\subset\W$ and 0~otherwise. This makes it possible 
to interpret the classical probability measure as an evolving ontological state. If a 
property is represented by a subspace~$\W$ of some vector space and the 
probability measure determined by a maximal measurement is represented by a 
(1-dimensional) ray~$\R$ in that space, probability~1 goes with $\R\subset\W$ 
and probability~0 goes with $\R\perp\W$, which makes room for nontrivial 
probabilities~\cite{justso,IUCAA}.]

\smallskip
(b) {\it Normalized\/} because as a possible outcome to which a probability is 
assigned, $\ket\psi$ is a shorthand notation for a projector $\ketbra\psi\psi$, 
and because the probabilities it assigns when it is an actual outcome add up to~1.

\smallskip
\item[(iii)] Why is a composite system ``described'' by the tensor product of the 
Hilbert spaces of the constituent systems? Given Born's rule, because
\be
p(a'|a)\,p(b'|b)=|\braket{a'}{a}\braket{b'}{b}|^2=
\left|\bigl[\bra{a'}\otimes\bra{b'}\bigr]
\bigl[\ket{a}\otimes\ket{b}\bigr]\right|^2.
\ee
\item[(iv)] Why do states ``evolve'' according to $i\hbar\ket{\dot 
\psi}=H\ket\psi$ with a Hermitian~$H$? Because according to Born's rule the 
probability of outcome $\ket{\psi_2}$ at~$t_2$ based on outcome 
$\ket{\psi_1}$ at~$t_1$ is given by
\be
|\sandwich{\psi_2(t_2)}{U(t_2,t_1)}{\psi_1(t_1)}|^2,
\ee
where $U$ is unitary. And why is $U$ unitary? Because probability is conserved 
whenever the system in question persists (is stable).
\end{enumerate}

Given that quantum states are probability measures, it stands to reason that they 
are $\sigma$-additive positive functionals on the projection operators in a Hilbert 
space~\cite{Jauch}. The usual choice of projection operators allows one to 
derive the trace rule,\cite{Gleason} and hence Born's rule, for Hilbert spaces of 
dimension $d\geq3$, without invoking~(iii), which makes it possible to 
understand the probabilistic origin of~(iii). By a perfectly justifiable 
generalization from projection valued measures to positive operator valued 
measures, the trace rule can be shown to hold for $d=2$ as well 
\cite{Fuchs,Busch,Cavesetal}. The reason why I prefer the Gleason-like 
derivations of Born's rule (given that quantum states {\it are\/} probability 
measures) to Zurek's is that the latter makes it so much harder (if not 
impossible) to justify assumptions (iii) and~(iv).

A final word: I do not wish to give the impression that the rejection of DEOQS is a 
cure-all. Rather, I believe that once our vision is no longer clouded by 
pseudoquestions entailed by the notion that quantum states are evolving 
ontological states, we will have a clearer view of the genuine issues, and a better 
chance of finding satisfactory solutions. For discussions of these issues see Refs. 
\cite{Pseudo}--\cite{IUCAA} and \cite{DQSE}--\cite{elusive}.

\end{document}